\documentstyle[12pt,epsfig]{article}
\begin{document}

\title{Perspectives of $e^+e^-$ production in
$pp$, $pd$ and $p Be$ reactions at SIS energies
\thanks{Work supported by GSI and BMBF} }
\author{E. L. Bratkovskaya, W. Cassing, and U. Mosel  \\[2mm]
{\normalsize Institut f\"{u}r Theoretische Physik, Universit\"{a}t Giessen}\\
{\normalsize 35392 Giessen, Germany}}
\date{}
\maketitle

\begin{abstract}
We study dilepton production from $pp$, $pd$ and $p Be$ collisions from
1 -- 5 GeV including the $\pi^0$, $\eta$, $\omega$ and $\Delta$ Dalitz
decays, direct decays of vector mesons ($\rho$, $\omega$) as well as
subthreshold $\rho$ production via baryonic resonances (e.g.
$D_{13}(1520), P_{11}(1710)$). Our calculations compare rather well
with the $pp$ and $pd$ data from the DLS Collaboration, however,
overestimate slightly the 'old' $p Be$ data from that group.
Futhermore, detailed predictions for differential dilepton spectra at
SIS energies are made with a high mass resolution that can be
controlled experimentally by the HADES Collaboration in near future.
\end{abstract}

\vspace{0.5cm}\noindent

\vspace{0.5cm}\noindent

\newpage
\section{Introduction}

Electromagnetic probes such as dileptons provide the most clear signals from
the early phases of high-energy heavy-ion collisions because they
may leave the reaction volume essentially undistorted by final-state
interactions.  Differential dilepton spectra from heavy-ion collisions
thus can provide information about the effective degrees of freedom at
high baryon density and temperature. Apart from that the in-medium
properties of hadrons (cf. Refs.
\cite{BrownRho,H&L92,Shakin94,Klingl96,Asakawa93,mosel91}), that are
closely related to the latter problem, can be explored as well.
According to QCD sum rules \cite{H&L92,Asakawa93,Leupold} as well as
QCD inspired effective Lagrangian models
\cite{BrownRho,Shakin94,Klingl96,Herrmann,asakawa,Chanfray,Rapp,Friman,RappNPA,Peters}
especially the vector mesons ($\rho$, $\omega$ and $\phi$)
significantly change their properties with nuclear density.  It has
been argued, that the experimentally observed enhanced production of
soft lepton pairs above known sources in nucleus-nucleus collisions at
SPS energies \cite{CERES,HELIOS} might be due to the in-medium
modification of vector mesons \cite{Li,Li96,Cass95CH,Brat97,CBRW97}
rather than reflecting a new state of hadronic matter.

Dileptons from heavy-ion collisions have also been measured by the DLS
Collaboration \cite{DLSold,DLSnew} at BEVALAC energies, where a
different temperature and density regime is probed.  The first
generation of DLS data \cite{DLSold}, based on a limited data set, were
consistent with the results from early transport model calculations
\cite{Xiong90,Wolf90,Gudima,BCMas96} including the conventional
dileptons sources as $pn$ bremsstrahlung, $\pi^0$, $\eta$, $\omega$ and
$\Delta$ Dalitz decay, direct decay of vector mesons and pion-pion
annihilation, without incorporating any medium effects.  A more recent
DLS measurement \cite{DLSnew} including the full data set and an
improved analysis, however, shows an increase by about a factor of 5-7
in the cross section in comparison to Ref. \cite{DLSold} and the early
theoretical predictions.

This enhancement of the $e^+e^-$ yield in nucleus-nucleus collisions
provides some theoretical 'puzzle'. In Ref. \cite{BCRW97} the in-medium
$\rho$ spectral functions from Refs.~\cite{RappNPA,Peters} have been
implemented in the HSD transport approach for $\rho$ mesons produced in
baryon-baryon, pion-baryon collisions as well as from $\pi\pi$
annihilation, and a factor of 2-3 enhancement has been obtained
compared to the case of a free $\rho$-spectral function. In
Ref.~\cite{Ernst} dropping vector meson masses and $\omega$ meson
broadening were incorporated in the UrQMD transport model, which also
showed an enhancement of the dilepton yield, however, the authors could
not describe the new DLS data \cite{DLSnew} as well.  Another attempt
to solve the DLS 'puzzle' has been performed in Ref.~\cite{BrKo98}
where the dropping hadron mass scenario was considered together with
the subthreshold $\rho$ production in $\pi N$ scattering via the
baryonic resonance $N(1520)$ whose importance was pointed out in
Refs.~\cite{Peters,Stony98}.  It was found that the enhancement of the
dilepton spectra due to low mass $\rho$'s from the $N(1520)$ was not
sufficient to match the DLS data.  Thus, all in-medium scenarios that
successfully have explained the dilepton enhancement at SPS energies
failed to describe the new DLS dilepton data \cite{DLSnew} from
heavy-ion collisions ($^{12}C + ^{12}C$ and $^{40}Ca + ^{40}Ca$).

In 1998 the DLS Collaboration has, furthermore, published dilepton
data from elementary $pp$ and $pd$ collisions at 1-5~GeV \cite{DLSpp}
which provides the possibility for an independent check of the
elementary dilepton channels that enter as 'input' in transport
calculations for heavy-ion reactions.  Such an analysis has been
carried out in Refs.~\cite{Ernst,Brat99} and it was shown that the
dilepton invariant mass spectra from $pp$ reactions in the energy
range from 1 -- 5 GeV could be rather well described within the
standard sources of hadronic Dalitz and direct decays when including
also the 'subthreshold' production of $\rho$-mesons by baryonic
resonances \footnote{For a detailed description of the individual
channels we refer the reader to Ref. \cite{Brat99}.}.  It has been,
furthermore, argued that additional channels (like dilepton decays of
scalar mesons ($a_0, f_0$) or four-body final states decays like
$\eta^\prime\to \pi^+\pi^-e^+e^-$, $(\rho^0,\omega)\to
\pi^+\pi^-e^+e^-$, $(\rho^0,\omega)\to \eta \pi^0 e^+e^-$) might
contribute to the elementary yield in $pp$ and $pd$ reactions
\cite{Tuebingen}.  Thus it is indispensible, that also these elementary
reactions have to be measured again with high resolution in order to
understand more accurately the $e^+e^-$ radiation mechanisms.

In this article we perform a detailed study of dilepton production from
$pp$, $pd$ and $p Be$ collisions from 1 -- 5 GeV including all channels
as described in Ref. \cite{Brat99}, i.e. the subthreshold $\rho$
production via baryonic resonances (e.g. $D_{13}(1520), P_{11}(1710)$)
as well as $\pi^0$, $\eta$, $\omega$ and $\Delta$ Dalitz decays and the
direct decay of vector mesons ($\rho$, $\omega$, $\phi$). We
investigate the role of isospin degrees of freedom with respect to
dilepton production in $pd$ reactions by employing exact
energy-momentum conservation for the deuteron target. The $p Be$
reactions are calculated within the resonance transport model
\cite{Effe99gam} that includes the same elementary production
amplitudes.

The paper is organized as follows:  In Section~2 we briefly present the
underlying resonance model that enters the coupled-channel BUU
transport approach.  In Section~3 we provide a comparison of our
results with the corresponding DLS data including the DLS acceptance
and resolution functions. In Section 4 we, furthermore, give detailed
predictions for these reactions with a high mass resolution in view of
upcoming experiments with the HADES detector at GSI Darmstadt. We close
with a summary and discussion of open problems in Section 5.

\section{Description of the model}

\subsection{Resonance approach}

We perform our analysis of dilepton production from $pp$, $pd$
and $pBe$ collisions within the resonance approach of
Refs. \cite{Effe99gam,EffePhD}.
This model is based on the resonance concept of nucleon-nucleon and
meson-nucleon interactions at low invariant energy $\sqrt{s} \ $
\cite{TeisZP97} by adopting all resonance parameters from the Manley
analysis \cite{Manley};
all states with at least 2 stars in Ref.~\cite{Manley}
are taken into account:
$P_{33}$(1232), $P_{11}$(1440), $D_{13}$(1520), $S_{11}$(1535),
$P_{33}$(1600), $S_{31}$(1620), $S_{11}$(1650), $D_{15}$(1675),
$F_{15}$(1680), $P_{13}$(1879), $S_{31}$(1900), $F_{35}$(1905),
$P_{31}$(1910), $D_{35}$(1930), $F_{37}$(1950), $F_{17}$(1990),
$G_{17}$(2190), $D_{35}$(2350). These resonances couple to the following
channels: $N \pi$, $N \eta$, $N \omega$, $\Lambda K$, $\Delta(1232)
\pi$, $N \rho$, $N \sigma$, $N(1440) \pi$, $\Delta(1232) \rho$
with respect to the production and decay.

It has been shown, that the resonance model provides a good description
of the experimental data on one- and two-pion production in
nucleon-nucleon collisions at low energy \cite{TeisZP97}. However, with
increasing energy the resonance contributions underestimate the data;
the missing yield is then treated as a backgraund term to the resonance
amplitude.  This background term 'mimics' $t$-channel particle
production mechanism as well as other non-resonance contributions (e.g.,
direct $NN\to NN\pi$, without creating an intermediate resonance).

With increasing energy, furthermore, the multiparticle production
becomes more and more important.
The high energy collisions -- above $\sqrt{s}$ = 2.6~GeV for
baryon-baryon collisions and $\sqrt{s}$ = 2.2~GeV for meson-baryon
collisions -- are described by  the LUND string fragmentation model
FRITIOF \cite{FRITIOF}. This aspect is similar to that used in the HSD
approach \cite{Ehehalt,Brat97,Geiss,CB99PR} and the UrQMD model
\cite{Bass}.

This combined resonance-string approach allows to calculate
particle production in baryon-baryon and meson-baryon
collisions from low to high energies.
The collisional dynamics for proton-nucleus reactions, furthermore, is
described by the coupled-channel BUU transport approach
\cite{Effe99gam,EffePhD} that is based on the same elementary cross
sections.

\subsection{Dilepton production}

The dilepton production within the resonance model can be schematically
presented in the following way:
\begin{eqnarray}
 BB &\to&R X   \label{chBBR} \\
 mB &\to&R X \label{chmBR} \\
      && R \to  e^+e^- X, \label{chRd} \\
      && R \to  m X, \ m\to e^+e^- X, \label{chRMd} \\
      && R \to  R^\prime X, \ R^\prime \to e^+e^- X. \label{chRprd}
\end{eqnarray}
In words: in a first step a resonance $R$ might be produced
in baryon-baryon ($BB$) or meson-baryon ($mB$) collisions -- (\ref{chBBR}),
(\ref{chmBR}). Then this resonance can couple to  dileptons
directly -- (\ref{chRd}) (e.g., Dalitz decay of the $\Delta$ resonance:
$\Delta \to e^+e^-N$) or decays to a meson $m$ (+ baryon) -- (\ref{chRMd})
which produces dileptons via direct decays ($\rho, \omega$)
or Dalitz decays ($\pi^0, \eta, \omega$).
The resonance $R$ might also decay into another resonance $R^\prime$ --
(\ref{chRprd}) which later produces dileptons via Dalitz decay or again
via meson decays  (e.g., $D_{35}(1930)\to\Delta\rho,\ \Delta\to
e^+e^-N, \ \rho\to e^+e^-$).  Note, that in our combined model the
final particles -- which couple to dileptons -- can be produced also
via non-resonant mechanisms, i.e. 'background' at low and intermediate
energies and string decay at high energies.

The electromagnetic part of all conventional dilepton sources  --
$\pi^0, \eta, \omega$ and $\Delta$ Dalitz decay, direct decay of vector
mesons $\rho, \omega$ -- are treated in the same way as described in
detail in Sec. 3 of Ref.~\cite{Brat99} where we have calculated the
dilepton spectra from $pp$ collisions in comparison to the DLS data.
The only difference is that in our present calculation we use the
spectral function of vector mesons ($\rho, \omega$) in the same form
as for all other resonances as used in Manley analysis \cite{Manley}:
\begin{equation}
f(M) =  N_V \ {2\over \pi} \ {M^2 \Gamma^V_{tot}
\over (M^2-m_V^2)^2 + (M {\Gamma_{tot}^V})^2}.
\label{fmdist}\end{equation}
In (\ref{fmdist}) $N_V$ guarantees normalization to unity, i.e. $\int
f(M) dM =1$, while $\Gamma_{tot}^V$ is the total vector meson width.

For $pd$ and $pBe$ collisions we also take into account $pn$
bremstrahlung using the soft-photon approximation \cite{Gale87}, where
the radiation from internal meson lines is neglected and the strong
interaction vertex is assumed to be on-shell.  In this case the strong
interaction part and the electromagnetic part separate; however, the
cross section for dileptons has to be corrected
\cite{GaleKap89,Schaefer89} by reducing the phase-space for the
colliding particles in their final-state.
We discard $pp$ bremsstrahlung from the nucleon pole since
the microscopic OBE calculations \cite{Schaefer,BrTit,Fred}
have shown that at 1.0 GeV the $pp$ bremsstrahlung is smaller than
the $\Delta$ Dalitz decay contribution by a factor of 2-3.
At this energy also the interference between nucleon and $\Delta$-pole
terms is negligible. At high energies, where this interference
becomes important \cite{Schaefer}, however, the overall contribution from
these channels is negligible (see below). We, furthermore, discard
$e^+e^-$ production from the scalar mesons $f_0, a_0$ as advocated in
Ref.~\cite{Tuebingen} since our studies in Ref.~\cite{Brat_Kon}
have indicated that the multiplicity of scalar mesons in $pp$ reactions
is very small in the energy regime of interest.

The basic underlying assumption for dilepton production, that enters in
our calculation, is that the amplitude can be factorized into a meson
or/and resonance production and a dilepton decay part.
This assumption provides not only a significant simplification, which
is necessary for applications in transport calculations
\cite{Wolf90,CB99PR,Effe99gam}, but also allows to take into account the
'inelasticity' from many-particle production channels which become
dominant at high energy.

\begin{figure}[t]
\phantom{a}\vspace*{2cm}
\centerline{\psfig{figure=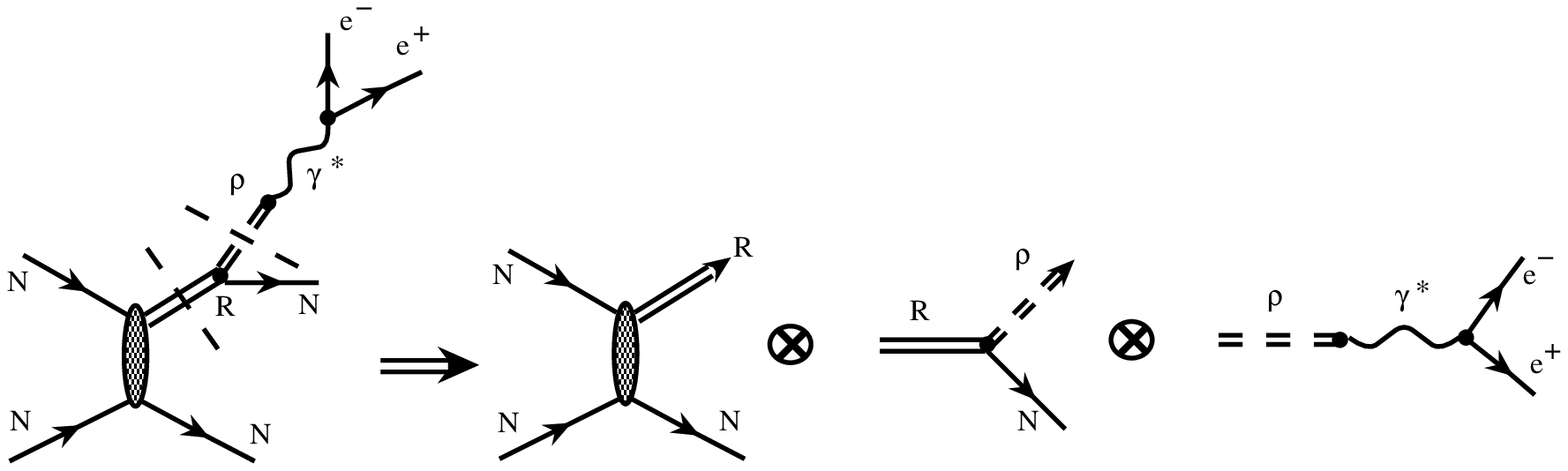,width=16cm}}
\phantom{a}\vspace*{2cm}
\caption{Factorization of the diagram for the process
$NN\to RN \to \rho NN \to e^+e^- NN$ with an intermediate
baryon resonance $R$.}
\label{Fig1SIS}
\end{figure}

In Fig.~\ref{Fig1SIS} we illustrate our factorization assumption for
the process $NN\to RN \to \rho NN \to e^+e^- NN$ with an intermediate
resonance $R$ coupled to the $\rho$ meson, e.g. $D_{13}(1520)$,
which is responsible for $\rho$ production below the threshold.
As shown in Fig.~\ref{Fig1SIS} we cut the diagram twice such that we
have a resonance production part ($NN\to RN$), a resonance decay to the
$\rho$ meson ($R\to \rho N$) and direct dilepton decay of the $\rho$ meson
($\rho\to e^+e^-$). This assumption with respect to subthreshold $\rho$
production via the $D_{13}(1520)$ is different to the ansatz used in
the previous dilepton analysis \cite{BrKo98,Brat99}. In Ref. \cite{Brat99}
we have factorized the amplitude $NN\to RN \to \rho NN \to e^+e^- NN$ into
a $D_{13}(1520)$ production part ($NN\to RN$) and Dalitz decay part
$R\to \rho N \to e^+e^-N$ (cf. Sect.  3.1 of Ref. \cite{Brat99}).
In the latter approach the $\rho$ meson is considered to be off-shell, i.e.
a virtual particle, and can have a mass below $2 m_\pi$ which leads to
the Dalitz behavior of the dilepton spectra (cf. Fig. 3 of Ref.
\cite{Brat99}), whereas in the present approach we assume that the $\rho$
meson has a free spectral function with masses only above $2m_\pi$.
The latter assumption leads to the 'direct $\rho$ decay' profile for
dileptons from subthreshold $\rho$'s and is consistent with our
resonance prescription used also in the transport calculations
\cite{Effe99gam}, where we propagate such $\rho$ mesons explicitly.

To compare with the experimental data of the DLS Collaboration one has
to use the appropriate experimental filter, which is a function of the
dilepton invariant mass $M$, transverse momentum $p_T$ and rapidity
$y_{lab}$ in the laboratory frame  -- $F(M,p_T,y_{lab})$. For that
purpose we explicitly simulate all dileptons by Monte-Carlo
and apply the filter $F(M,p_T,y_{lab})$.

\section{$e^+e^-$ results in comparison to DLS data}

\subsection{Dileptons from $pp$ collisions}

In Fig.~\ref{Fig2SIS} we present the calculated dilepton invariant mass
spectra $d\sigma/dM$ for $pp$ collisions from 1.0 -- 4.9 GeV including
the final mass resolution and filter $F(M,p_T,y_{lab})$ from the DLS
Collaboration (version 4.1) in comparison to the DLS data \cite{DLSpp}.
The thin lines indicate the individual contributions from the different
production channels; {\it i.e.}~ starting from low $M$:
Dalitz decay $\pi^0 \to \gamma e^+ e^-$ (short dashed line),
$\eta \to \gamma e^+ e^-$ (dotted line),
$\Delta \to N e^+ e^-$ (dashed line),
$\omega \to \pi^0 e^+ e^-$ (dot-dashed line),
for $M \approx $ 0.7 GeV: $\omega \to e^+e^-$ (dot-dashed line),
$\rho^0 \to e^+e^-$ (short dashed line).
The full solid line represents the sum of all sources considered here.

Whereas at 1.04 GeV the dileptons stem practically all from $\pi^0$ and
$\Delta$ Dalitz decays, $\rho$ decay becomes more important  with
increasing energy and at high invariant mass.
Note, that in the present analysis we do not separate the production
mechanism for $\rho$ mesons as in Ref.~\cite{Brat99}, i.e. the
$\rho$-meson contribution might come from the decay of $D_{13}(1520)$
and $P_{11}(1710)$ or non-resonance $\rho$ production. Thus, the short
dashed line (denoted as $\rho^0 \to e^+e^-$) presents the sum of all
$\rho$ contributions.  However, the resonance
mechanism dominates for $\rho$ production at low energies, especially
below  the $pp\to \rho pp$ threshold $\sqrt s < 2m_N+m_\rho$, where
$\sqrt s$ is the invariant collision energy (see Sec. 2 in
Ref.~\cite{Brat99}).

At 2.1 and 4.9 GeV the dilepton yield is dominated by the $\eta$ Dalitz
decay and direct decays of $\rho$ and $\omega$ mesons.  The $\rho$
spectrum is enhanced towards low $M$ due to the limited phase space and
strong mass dependence $(M^{-3})$ of the dilepton decay width
\cite{Brat99} in line with the vector dominance model.
Note, that at 2.1 GeV we use the resonance production mechanism in
our model, whereas at 4.9 GeV most of the particles are produced by
string decays.  Within the strict resonance model which gives
only the exclusive $pp\to pp\eta$ cross section the $\eta$ yield from
$pp$ collisions is underestimated by about a factor of 2
(dot-dot-dashed line) at 2.1 GeV. Since at this energy other
channels such as $NN\eta\pi, NN\eta\pi\pi$ are open, we have
estimated the cross section for these channels by the string model and
added it to the total $\eta$ yield.  This gives the final $\eta$
contribution (dotted line) for 2.1 GeV.

\subsection{Dileptons from $pd$ collisions}

The experimental measurement of dilepton spectra from proton and
deuteron targets provides a unique possibility to get information about
the isospin dependence of the $NN$ interaction which is interesting in
itself and, furthermore, important with respect to heavy-ion reactions.
The $pp$ and $pd$ dilepton data from the DLS Collaboration \cite{DLSpp}
are the first experimental step in this direction.

Since the deuteron is a bound system of proton and neutron, the Fermi
motion of the nucleons plays an important role for all interactions. For
example, Fermi motion leads to a shift of the threshold for
particle production in $pd$ interactions to lower $\sqrt{s}$, i.e. the
channels closed by kinematics in $pp$ collisions might be
open in $pd$. Thus, the  $pd$ reactions can not be directly approximated
by the simple sum of $pp$ and $pn$.  On the other hand, one can't
treat the deuteron within the usual mean-field approach used in transport
calculations for more heavy nuclei.

In our analysis we thus take into account the Fermi motion of nucleons using
the momentum distribution of the deuteron calculated with
the Paris potential \cite{Paris}.
We initialize a deuteron in momentum space (using a Monte Carlo method)
according to the momentum distribution $f(p)$:
\begin{eqnarray}
f(p) = {|\Psi(p)|^2\over |\Psi(0)|^2},
\label{deut}\end{eqnarray}
where $p$ is the momentum of a nucleon in the deuteron center-of-mass,
while $\Psi(p)$ is the deuteron wave function.
To fulfill the binding energy constraint we assume
\begin{eqnarray}
E_{spec}=\sqrt{p^2+m_N^2}, \ \ E_{part}=E_d-E_{spec}.
\label{disper}\end{eqnarray}
The nucleon, which does not participate in the reaction, i.e. the spectator,
has a free dispersion relation for the energy-momentum $E_{spec}$,
whereas $E_{part}$ is the energy of the nucleon participating
in the reaction (participant).
In (\ref{disper}) $E_d = 2m_N+\varepsilon$ is the total deuteron energy,
where $\varepsilon=-2.2$ MeV is the binding energy of the deuteron.

The invariant collision-energy distribution $dN/d\sqrt s$ for $pd$
reactions calculated with the dispersion relation (\ref{disper}) is
shown by the solid histogram in Fig.~\ref{Fig3SIS}.  For comparison
(dashed histogram) we present the $dN/d\sqrt s$ distribution in the
physical region ($2m_N \le \sqrt s  \le \sqrt{s_{max}}$) calculated
with $f(p)$ (\ref{deut}) in the quasi-free scattering limit, i.e. using
the free dispersion relation for both nucleons in the deuteron:
$E_{spec}=E_{part}=\sqrt{p^2+m_N^2}$.
The lower limit of $\sqrt s$ is given by the free cross section that
only exists for $\sqrt \ge 2m_N$.
As seen from Fig.~\ref{Fig3SIS}, both prescriptions are the same in the
peak value, however, the deuteron spectral distribution (\ref{disper})
leads to an enhancement of the $\sqrt s$ distribution at small $\sqrt
s$ and a reduction at high $\sqrt s$. This implies a substantial
lowering of particle production at high $\sqrt s$ with respect to the
quasi-free limit, which has been often used in the literature.

\begin{figure}[h]
\centerline{\psfig{figure=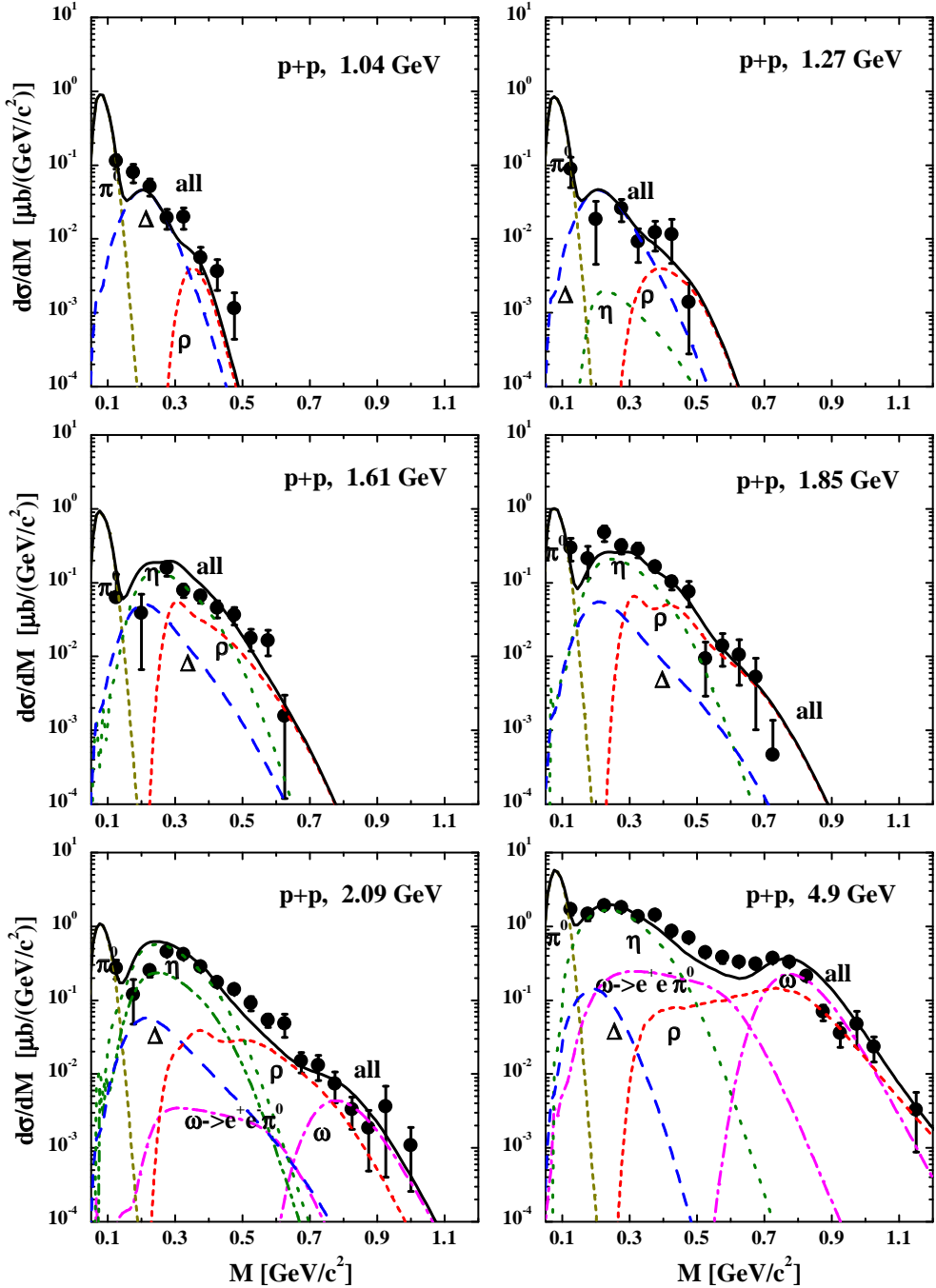,width=13cm}}
\caption{The calculated dilepton invariant mass spectra $d\sigma/dM$
for $pp$ collisions from 1.0 -- 4.9 GeV in comparison to the DLS data
\protect\cite{DLSpp}.  The thin lines indicate the individual
contributions from the different production channels; {\it i.e.}~
starting from low $M$:  Dalitz decay $\pi^0 \to \gamma e^+ e^-$ (short
dashed line), $\eta \to \gamma e^+ e^-$ (dotted line), $\Delta \to N
e^+ e^-$ (dashed line), $\omega \to \pi^0 e^+ e^-$ (dot-dashed
line); for $M \approx $ 0.7 GeV: $\omega \to e^+e^-$
(dot-dashed line), $\rho^0 \to e^+e^-$ (short dashed line).  The
full solid line represents the sum of all sources considered here.
At 2.09 GeV the dot-dot-dashed line indicates the $\eta$ contribution from
the resonance decays only while the dotted line corresponds to
the inclusive $\eta$ production (see text).}
\label{Fig2SIS}
\end{figure}

\begin{figure}[t]
\phantom{a}\vspace*{-2cm}
\centerline{\psfig{figure=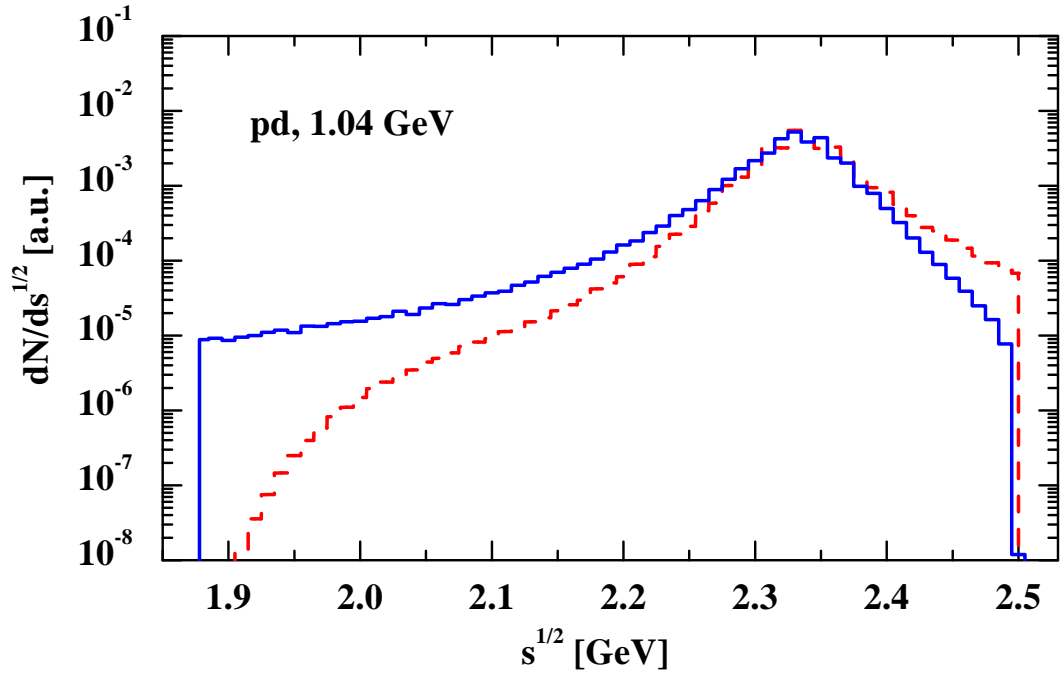,width=14cm}}
\caption{The distribution in the invariant collision energy $\sqrt{s}$
($dN/d\sqrt s$) in arbitrary units for the $pd$ reaction at 1.04 GeV
calculated with the dispersion relation (\protect\ref{disper})
(solid histogram) and within the quasi-free scattering limit
(dashed histogram).}
\label{Fig3SIS}
\end{figure}

Within the 'binding' prescription (\ref{disper}) we have calculated the
dilepton invariant mass spectra $d\sigma/dM$ for $pd$ collisions from
1.0 -- 4.9 GeV including the final mass resolution and
filter $F(M,p_T,y_{lab})$ from the DLS Collaboration (version 4.1).
The results are show in Fig.~\ref{Fig4SIS} in comparison to the DLS data
\cite{DLSpp}, where the assignment of the lines is the same as in
Fig.~\ref{Fig2SIS}. Additionally, the dot-dot-dashed lines indicate the
$pn$ bremsstrahlung contributions.

Similar to Fig.~\ref{Fig2SIS}, at 1.04 GeV the $\pi^0$ and $\Delta$
dalitz decays are dominant, however, due to the larger mass of the
deuteron, the $\eta$ channel is already open contrary to the $pp$
reaction and the subthreshold $\rho$ branch is getting more important
at high invariant mass $M$. Also $pn$ bremsstrahlung contributes at low
energies, however, with increasing bombarding energy it becomes less
important.

Starting from 1.61 GeV the dominant contribution at low $M$ is again
the $\eta$ Dalitz decay, whereas at high $M$ the dileptons stem
basically from direct $\rho, \omega$ decays.
The $\eta$ contribution here is calculated by assuming
$\sigma_{pn\to pn\eta}=6\sigma_{pp\to pp\eta}$ for
$2.425\ {\rm GeV} \le \sqrt s \le 2.517\ {\rm GeV}$ in line with
the experimental data from Ref.  \cite{Pinot},
$\sigma_{pn\to pn\eta}=2.5\sigma_{pp\to pp\eta}$ for
$2.517\ {\rm GeV} \le \sqrt s \le 2.65\ {\rm GeV}$ and
$\sigma_{pn\to pn\eta}\simeq \sigma_{pp\to pp\eta}$ for
$\sqrt s \ge 2.65\ {\rm GeV}$ where the multiparticle production
(above 2.65 GeV) is evaluated within the LUND string model. Note,
that at high $\sqrt s$ the isospin asymmetry $\sigma_{pn}/\sigma_{pp}$
should be washed out by the isospin of pions that are created together
with the $\eta$. Though this 'recipe' seems to work in comparison to
the DLS data, we point out that further experimental data with good
mass resolution will be necessery to clarify the situation for the
isospin couplings.

In order to demonstrate the importance of isospin degrees-of-freedom for
dilepton production we present in Fig.~\ref{Fig5SIS} the invariant
mass spectra $d\sigma/dM$ for $pp$ (short dashed line), $pn$ (long dashed
line), $pp+pn$ (dot-dashed line) and $pd$ (solid line) collisions at
1.61 GeV in comparison to the DLS data \cite{DLSpp}.
At these energies the $pn$ channel gives the main
contribution to the $pd$ dilepton yield. Furthermore, the influence of
the Fermi motion in the deuteron is clearly seen at high invariant mass $M$
by comparing the dot-dashed ($pp+pn$) and solid ($pd$) lines.

In Fig. \ref{Fig6SIS} we present the excitation function, i.e. the
integrated dilepton cross section for masses above 0.15 GeV,
as a function of the available energy $Q=\sqrt{s}-2m_p$ for $pp$
(dashed line) and $pd$ (solid line) in comparison to the DLS data
\cite{DLSpp} (open circles for $pp$ and full circles for $pd$).
The calculated excitation functions for $pp$ and $pd$ increase
with the energy and stay  within the experimental errorbars.

\subsection{Dileptons from $pBe$ collisions}

Since $Be$ is one of the lightest nuclei, in-medium effects play
a minor role because most hadron decays occur already in vacuum.
So one can consider $pBe$ collisions as an additional
and independent check for the elementary reactions and isospin degrees
of freedom. However, the collisional dynamics has to be taken into
account properly. For that purpose we use the coupled-channel BUU
transport approach as discussed in Sec. 2.

Dileptons from $pBe$ have been measured by the DLS Collaboration in the
end of the 80's \cite{DLSold}.  In Fig.~\ref{Fig7SIS} we show the
calculated dilepton invariant mass spectra $d\sigma/dM$ for $pBe$
collisions at 1.0 (upper part), 2.1 (middle part) and  4.9 GeV (lower
part) including the final mass resolution and filter
$F(M,p_T,y_{lab})$ from the DLS Collaboration (version 1.6) in
comparison to the data \cite{DLSold}. The assignment of the lines is
the same as in Figs.~\ref{Fig2SIS} and \ref{Fig4SIS}. With the filter
1.6 we do not get enough suppression for the dilepton yield at $M\le
0.15$ GeV which basically stems from $\pi^0$ Dalitz decay. Similar to
$pd$ reactions the dominant channels at 1.0 GeV are the $\Delta$ Dalitz
decay and $pn$ bremsstrahlung. At 2.1 and 4.9 GeV the $\eta$ Dalitz
decay becomes important at low $M$ and the direct decays of vector
mesons ($\rho, \omega$) for $M\sim 0.8$ GeV.  Within the present model
we slightly overestimate the data at 4.9 GeV, however, we discard any
discussion about the physical reason since the 'old' DLS filter is not
sufficiently understood.  For example, the 'old' DLS data for $pBe$ at
4.9 GeV show a similar dilepton yield as for $pd$, whereas according to
our calculations it should be higher by a factor of $\sim 2$ since more
nucleons are involved in the $pBe$ reaction.

In this respect we have performed  detailed calculations for $pp$, $pd$
and $pBe$ reactions without implementing any filter, however, including
a high (10 MeV) mass resolution for $d\sigma/dM$.  For the results,
that can be controlled by the HADES Collaboration in near future, we
refer to the following Section.

\section{Predictions for HADES}

In this Section we present our predictions for the dilepton invariant
mass spectra, transverse momentum ($p_T$) distributions and rapidity
($y$) distributions for $pp$, $pd$ and $pBe$ collisons at 1.5 GeV and
4.0 GeV, which will be measured with the HADES spectrometer at GSI.

\begin{figure}[t]
\phantom{a}\vspace*{-2cm}
\centerline{\psfig{figure=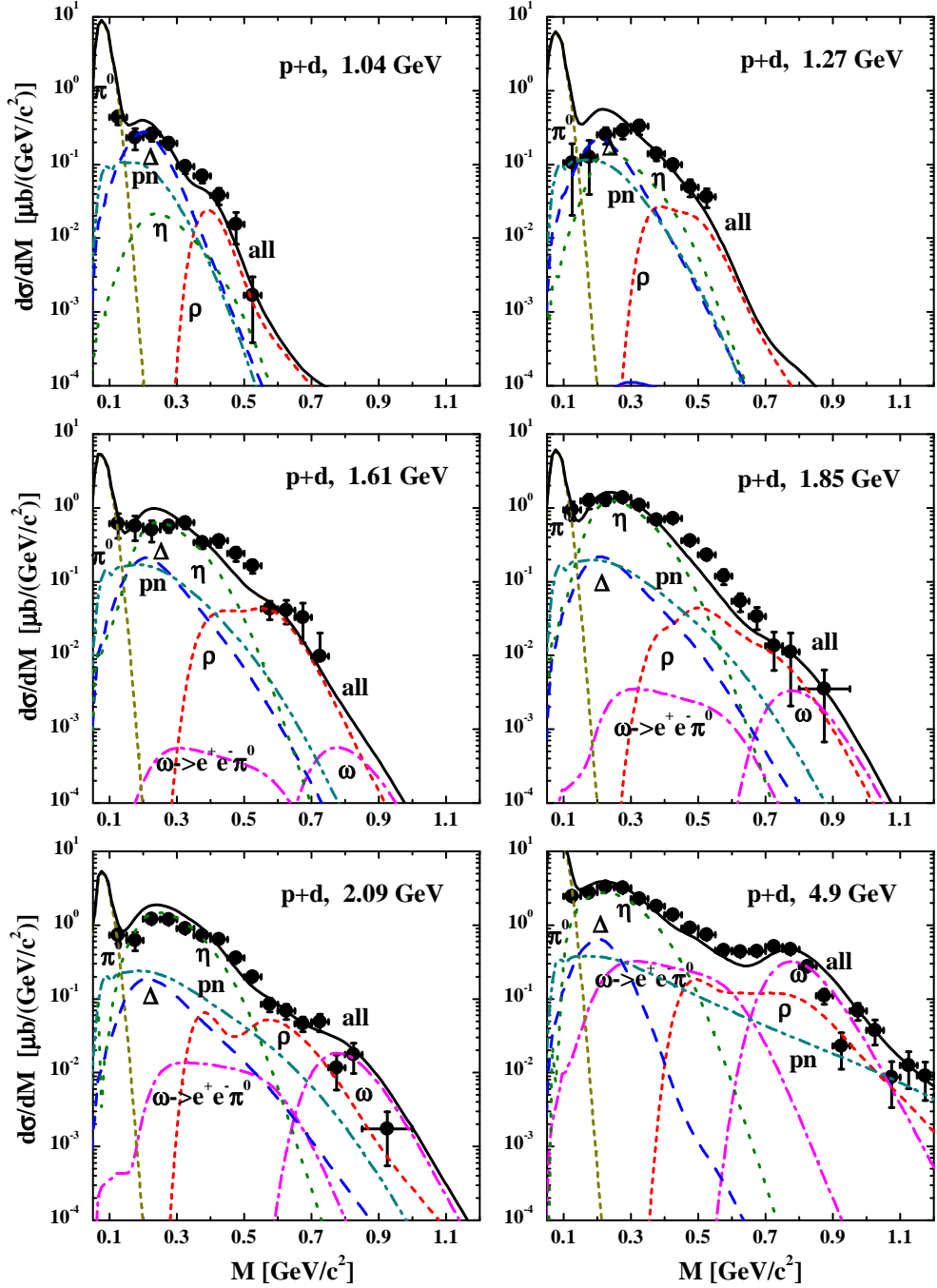,width=13cm}}
\caption{The dilepton invariant mass spectra $d\sigma/dM$ for
$pd$ collisions from 1.0 -- 4.9 GeV in comparison to the DLS data
\protect\cite{DLSpp}. The assignment of the lines is the same as in
Fig.~\protect\ref{Fig2SIS}. Additionally, the dot-dot-dashed lines
indicate the $pn$ bremstrahlung contributions.}
\label{Fig4SIS}
\end{figure}

\begin{figure}[t]
\centerline{\psfig{figure=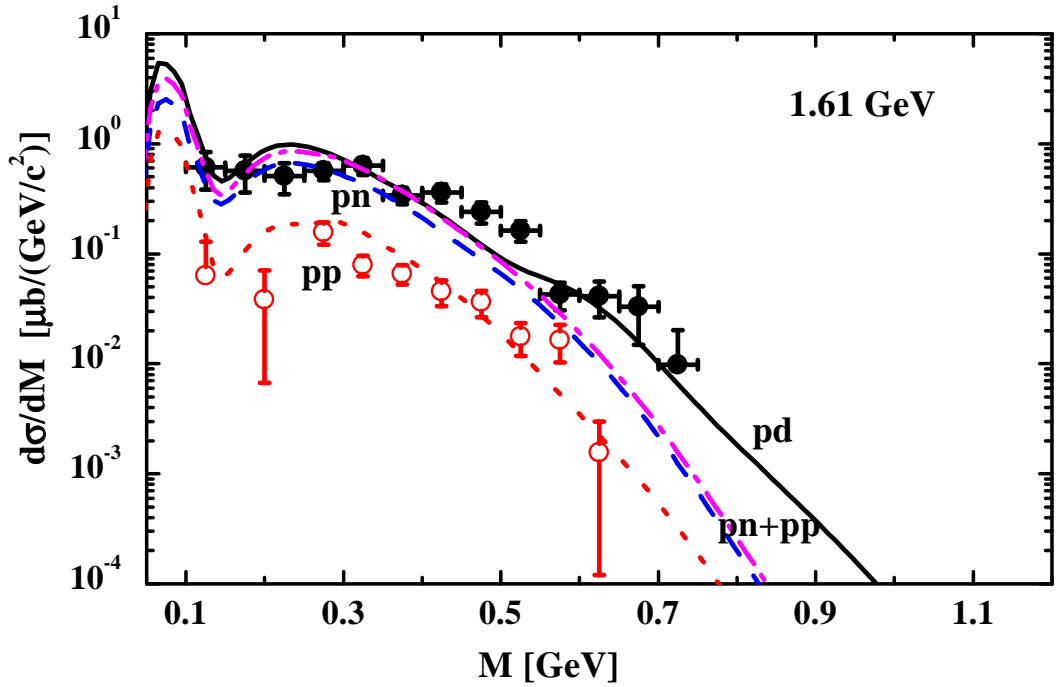,width=14cm}}
\caption{The dilepton invariant mass spectra $d\sigma/dM$ for
$pp$ (short dashed line),  $pn$ (long dashed line), $pp+pn$ (dot-dashed
line)  and $pd$ (solid line) collisions at 1.61 GeV in comparison
to the DLS data \protect\cite{DLSpp}. The difference between the
dot-dashed line and the solid line stems from the Fermi motion of
the nucleons in the deuteron.}
\label{Fig5SIS}
\end{figure}

\begin{figure}[t]
\centerline{\psfig{figure=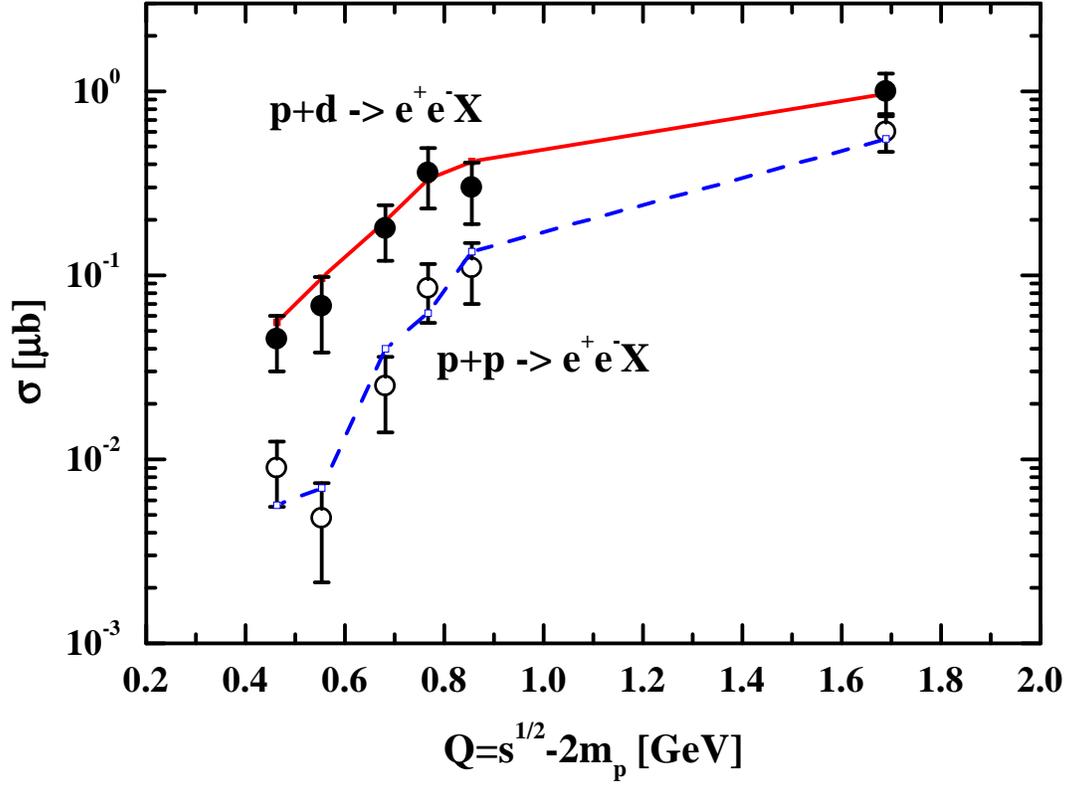,width=14cm}}
\caption{The excitation function for dileptons from $pp$
(dashed line) and $pd$ (solid line) in comparison to the DLS data
\cite{DLSpp} (open circles for $pp$ and full circles for $pd$ reactions).}
\label{Fig6SIS}
\end{figure}

\begin{figure}[t]
\phantom{a}\vspace*{-2cm}
\centerline{\psfig{figure=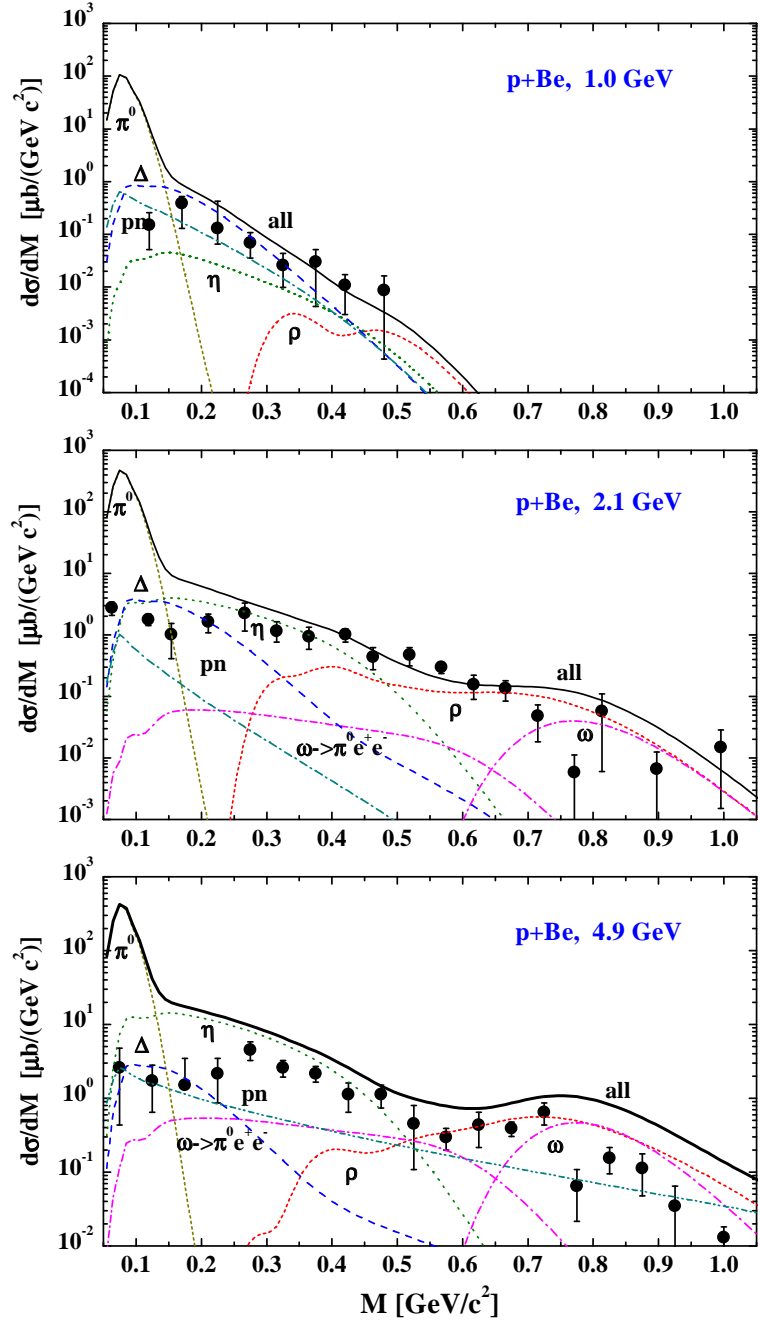,width=10cm}}
\caption{The dilepton invariant mass spectra $d\sigma/dM$ for $pBe$
collisions at 1.0 GeV (upper part), 2.1 GeV (middle part) and  4.9 GeV
(lower part) in comparison to the DLS data \cite{DLSold}.
The assignment of the individual lines is the same as in
Figs.~\protect\ref{Fig2SIS}, \protect\ref{Fig4SIS}.}
\label{Fig7SIS}
\end{figure}

\begin{figure}[t]
\phantom{a}\vspace*{-2cm}
\centerline{\psfig{figure=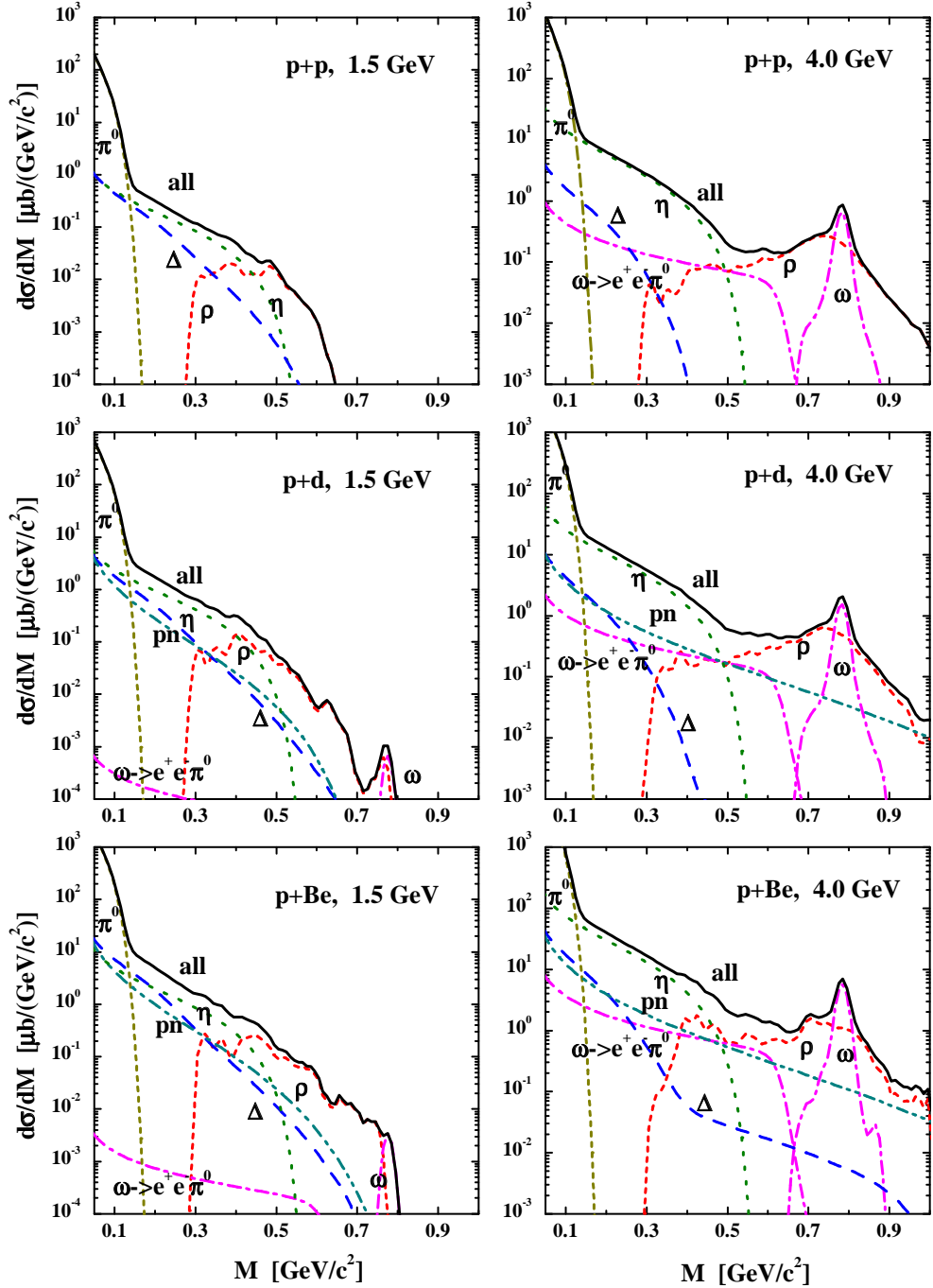,width=13cm}}
\caption{The dilepton invariant mass spectra $d\sigma/dM$ for
$pp$ (upper part), $pd$ (middle part) and $pBe$ collisions (lower part)
at 1.5 GeV (left panel) and 4.0 GeV (right panel) including a 10 MeV
mass resolution.  The assignment of the individual lines is  the same as
in Figs.~\protect\ref{Fig2SIS}, \protect\ref{Fig4SIS}.}
\label{Fig8SIS}
\end{figure}

In Fig.~\ref{Fig8SIS} the calculated dilepton invariant mass spectra
$d\sigma/dM$ for $pp$ (upper part), $pd$ (middle part) and $pBe$
collisions (lower part) at 1.5 GeV (left panel) and 4.0 GeV (right
panel) are shown including a $\Delta M=10$~MeV mass resolution.  The
assignment of the lines is the same as in Figs.~\ref{Fig2SIS} and
\ref{Fig4SIS}.  The dominant contribution at low $M$ ($> m_{\pi^0}$) is
the $\eta$ Dalitz decay, however, for $M > 0.4$ GeV the dileptons stem
basically all from direct vector meson decays ($\rho$ and $\omega$).
Within a 10 MeV mass resolution the peak from the direct $\omega$ decay
(which was completely smeared out within the DLS mass resolution) is
seen. The $\omega$ peak is getting more pronounced at high energy (cf.
the right panel of Fig.~\ref{Fig8SIS}).

\begin{figure}[t]
\phantom{a}\vspace*{-2cm}
\centerline{\psfig{figure=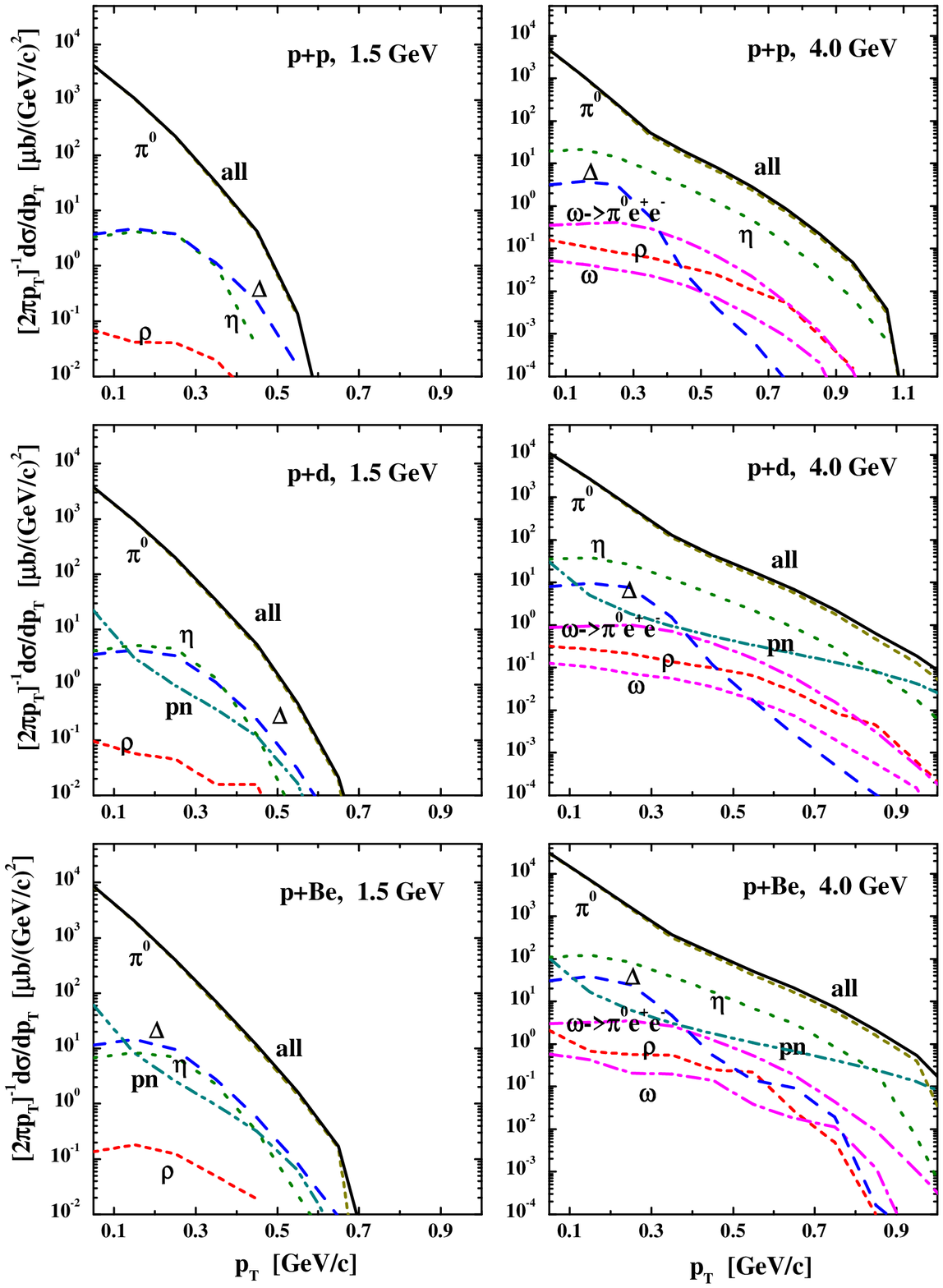,width=13cm}}
\caption{The transverse momentum distribution $d\sigma/dp_T /(2\pi p_T)$
for $pp$ (upper part), $pd$ (middle part) and $pBe$ collisions (lower
part) at 1.5 GeV (left panel) and 4.0 GeV (right panel).
The assignment of the individual lines is the same as
in Figs.~\protect\ref{Fig2SIS}, \protect\ref{Fig4SIS}.}
\label{Fig9SIS}
\end{figure}

In Fig.~\ref{Fig9SIS} we present the transverse momentum distribution
$d\sigma/dp_T /(2\pi p_T)$ for $pp$ (upper part), $pd$ (middle part)
and $pBe$ collisions (lower part) at 1.5 GeV (left panel) and 4.0 GeV
(right panel).  The assignment of the lines again is the same as in
Figs.~\ref{Fig2SIS} and \ref{Fig4SIS}.
Here we did not imply any cut in the invariant mass $M$. For all
systems and energies considered the $\pi^0$ Dalitz decay is much
higher than all other contributions, which is in line with the result of
Fig.~\ref{Fig8SIS} because the pion production cross section
is much larger than that for the other mesons. The shape of the $\pi^0,
\eta, \Delta, \omega$ Dalitz decays is defined by the kinematics of the
Dalitz decay and the 'flat' part at low $p_T$ depends on the mass of
the decaying resonance.

\begin{figure}[t]
\phantom{a}\vspace*{-2cm}
\centerline{\psfig{figure=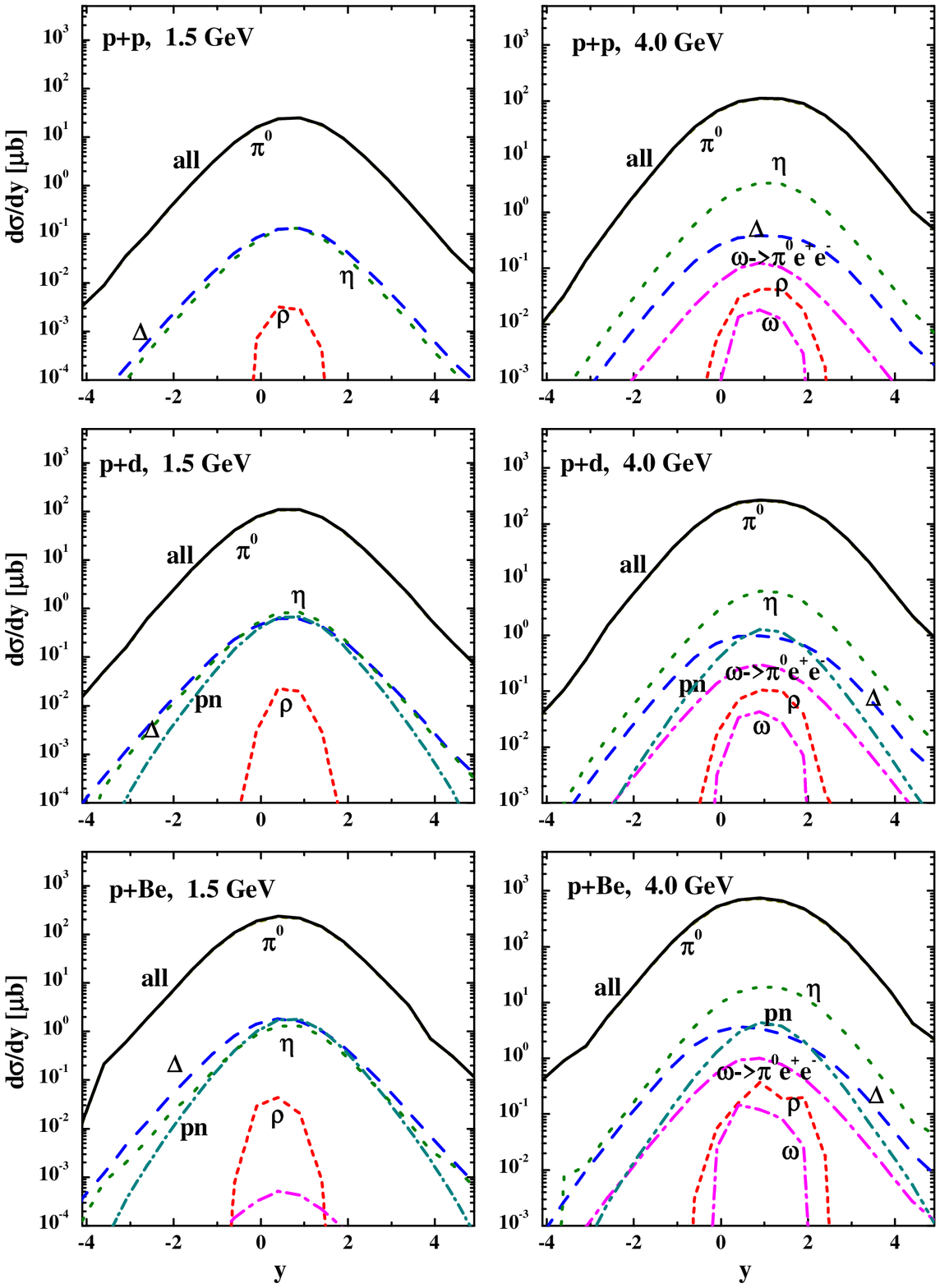,width=13cm}}
\caption{The rapidity distributions $d\sigma/dy$ for $pp$ (upper part),
$pd$ (middle part) and $pBe$ collisions (lower part) at 1.5 GeV
(left panel) and 4.0 GeV (right panel).
The assignment of the individual lines is the same as in
Figs.~\protect\ref{Fig2SIS}, \protect\ref{Fig4SIS}.}
\label{Fig10SIS}
\end{figure}

The laboratory rapidity distributions $d\sigma/dy$ for $pp$ (upper
part), $pd$ (middle part) and $pBe$ collisions (lower part) at 1.5 GeV
(left panel) and 4.0 GeV (right panel) are shown in
Fig.~\ref{Fig10SIS}, where the assignment of the lines is the same as
before.  The situation is similar to Fig.~\ref{Fig9SIS}, -- the $\pi^0$
Dalitz decay is the only visible contribution when imposing no cuts
on the invariant mass.

\begin{figure}[t]
\phantom{a}\vspace*{-2cm}
\centerline{\psfig{figure=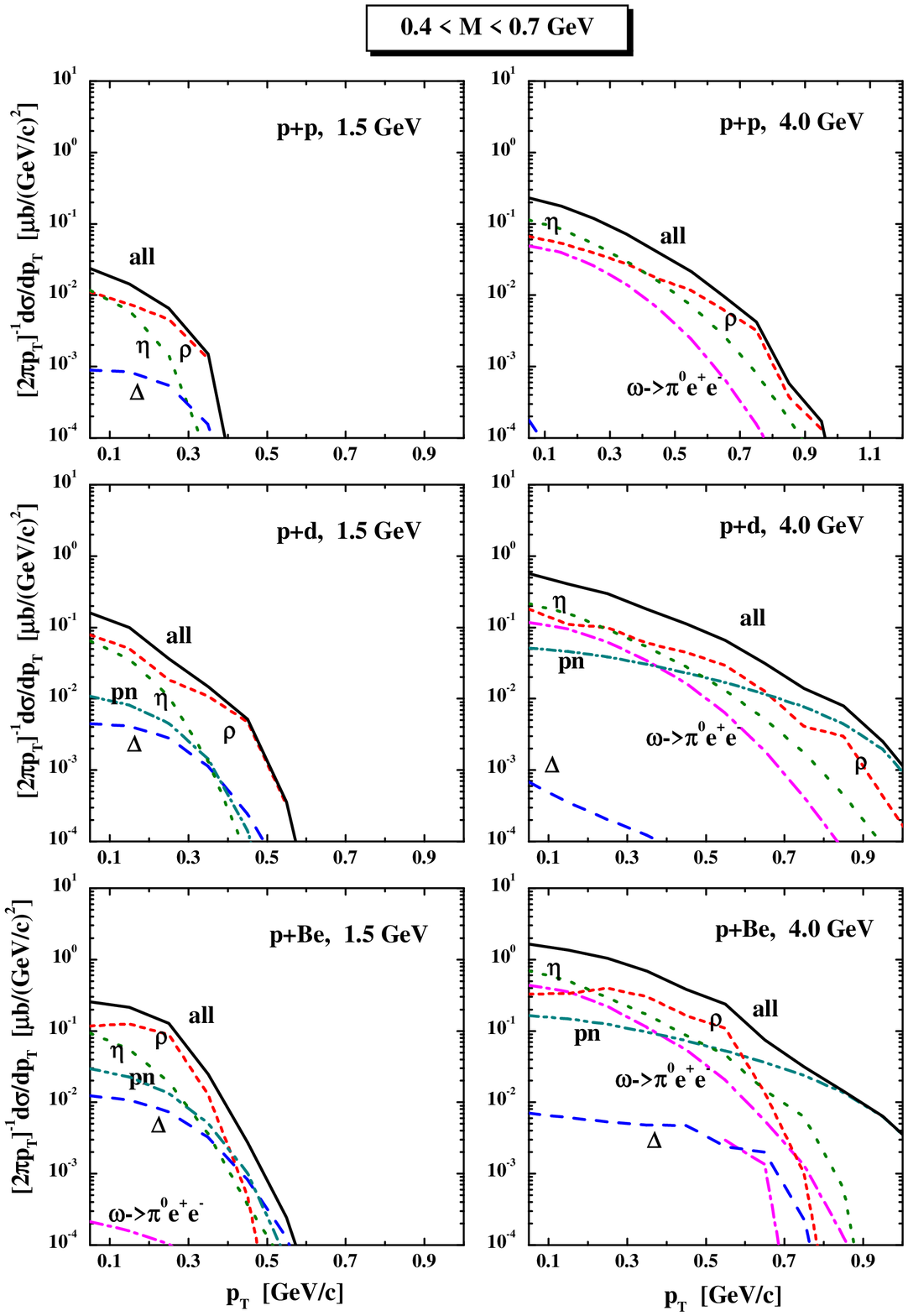,width=13cm}}
\caption{The transverse momentum distribution $d\sigma/dp_T /(2\pi p_T)$
for $pp$ (upper part), $pd$ (middle part) and $pBe$ collisions
(lower part) at 1.5 GeV (left panel) and 4.0 GeV (right panel)
implying a cut in invariant mass of $0.4 \le M \le 0.7$ GeV.
The assignment of the individual lines is the same as in
Figs.~\protect\ref{Fig2SIS}, \protect\ref{Fig4SIS}.}
\label{Fig11SIS}
\end{figure}

\begin{figure}[t]
\phantom{a}\vspace*{-2cm}
\centerline{\psfig{figure=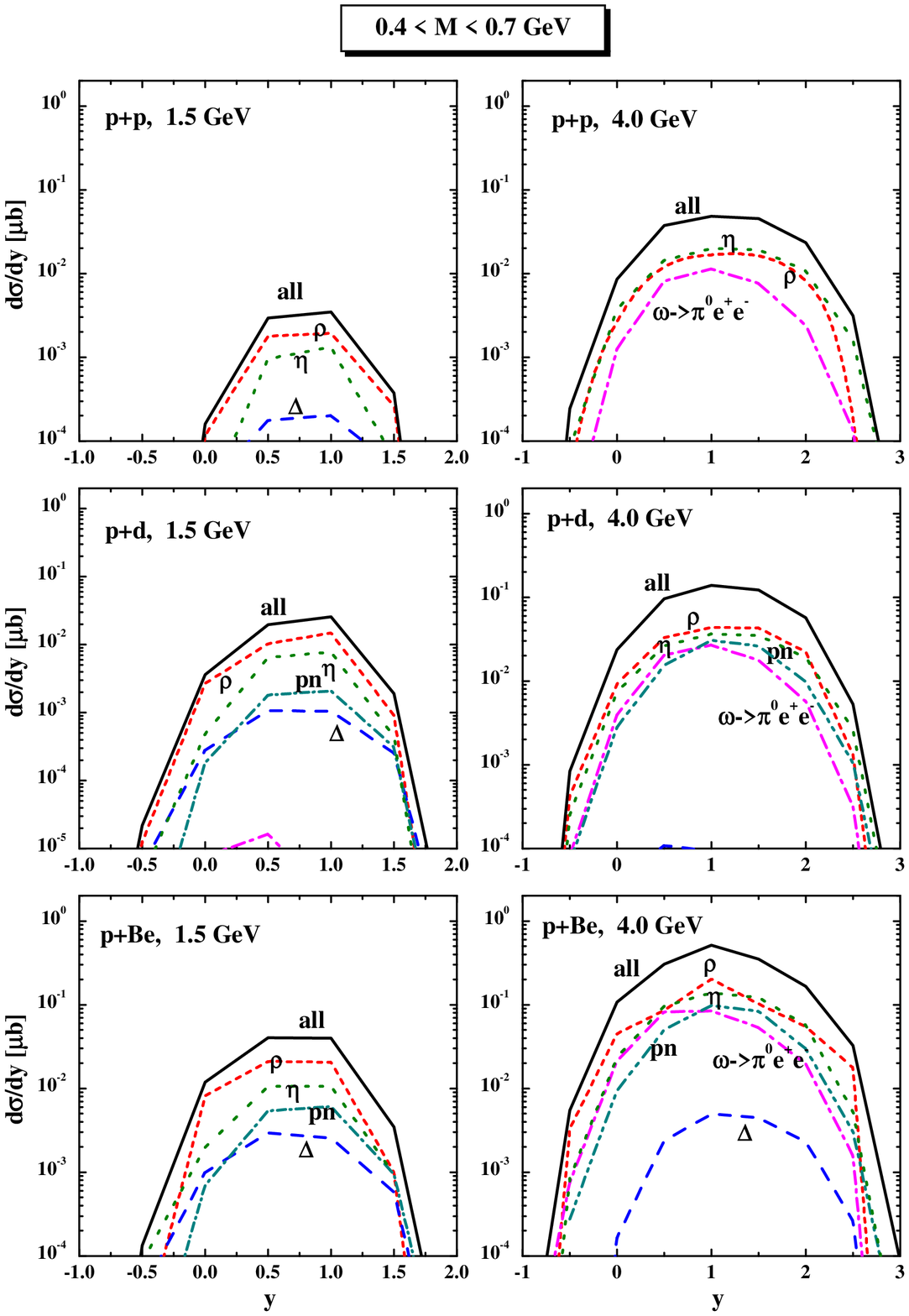,width=13cm}}
\caption{The laboratory rapidity distributions $d\sigma/dy$ for $pp$
(upper part), $pd$ (middle part) and $pBe$ collisions (lower part)
at 1.5 GeV (left panel) and 4.0 GeV (right panel)
implying a cut in invariant mass of $0.4 \le M \le 0.7$ GeV.
The assignment of the individual lines is the same as in
Figs.~\protect\ref{Fig2SIS}, \protect\ref{Fig4SIS}.}
\label{Fig12SIS}
\end{figure}

In order to make the contribution from other channels visible one has
to suppress the 'background' $\pi^0$ yield. In Fig.~\ref{Fig11SIS} we
display the transverse momentum spectra for $pp$ (upper part), $pd$
(middle part) and $pBe$ collisions (lower part) at 1.5 GeV (left panel)
and 4.0 GeV (right panel) using a cut in invariant mass as 0.4~GeV$\le
M \le 0.7$~GeV. Fig.~\ref{Fig12SIS} shows the laboratory rapidity
distributions for the same systems and kinematical conditions. With
this cut on intermediate dilepton masses one can select the region
where the $\pi^0$ Dalitz decay is totally removed and the $\eta$ Dalitz
decay is suppressed substantially (cf. Fig.~\ref{Fig8SIS}); in this way
the $\rho$ branch in the transverse momentum and rapidity spectra
becomes seen (cf. Figs.~\ref{Fig10SIS}, \ref{Fig11SIS}). Thus, a
measurement of the transverse momentum and rapidity spectra with a
corresponging cut in $M$ will provide additional and independent
information about the $\rho$ contribution to the dilepton spectra.

\begin{figure}[t]
\phantom{a}\vspace*{-2cm}
\centerline{\psfig{figure=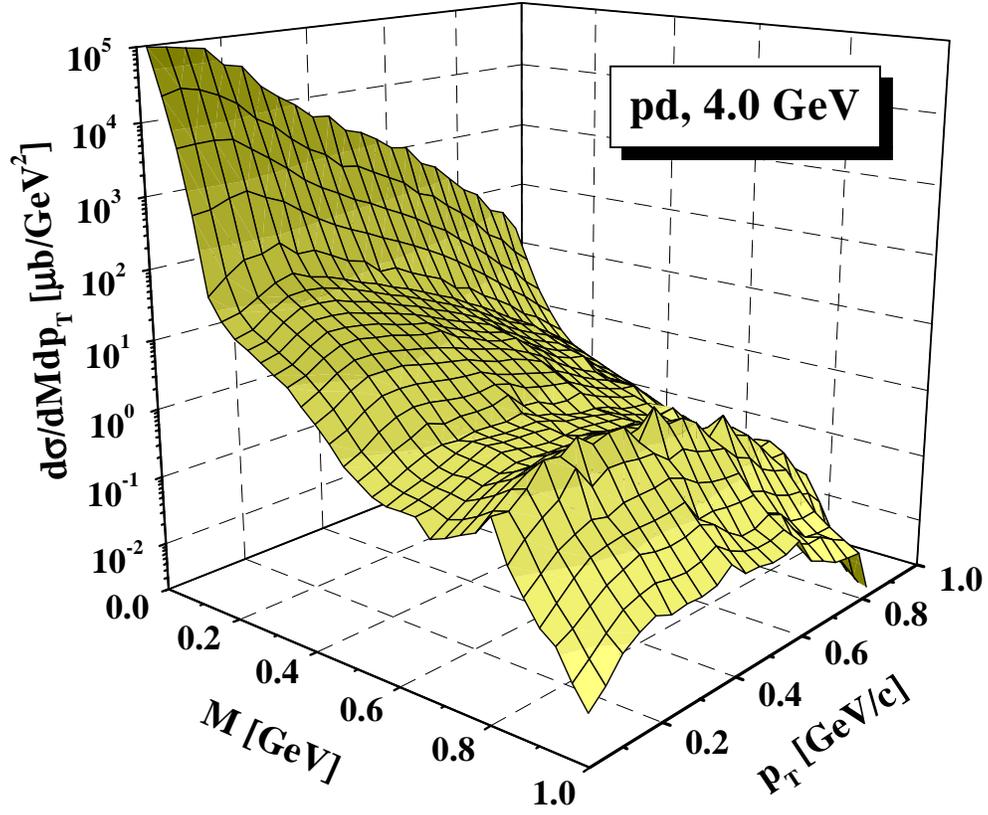,width=13cm}}
\caption{The calculated double differential dilepton spectra
$d\sigma/dMdp_T$ as a function of invariant mass $M$ and transverse
momentum $p_T$ for $pd$ collisions at 4.0 GeV.}
\label{Fig13SIS}
\end{figure}

Finally, in Fig.~\ref{Fig13SIS} we present the double differential
dilepton spectra $d\sigma/dMdp_T$ as a function of the invariant mass
$M$ and transverse momentum $p_T$ for $pd$ collisions at 4.0 GeV.
At fixed $p_T$ one can recognize the shape of the invariant mass
spectra (cf. Fig.~\ref{Fig8SIS}) with a strong $\pi^0$ Dalitz decay
branch at low $M$ as well as the contributions from $\eta$ Dalitz decay
and the vector meson ($\rho, \omega$) decays. At fixed $M$ the shape looks
similar to the one in Fig.~\ref{Fig9SIS}. At low $M$ the
exponential decrease stems from $\pi^0$ Dalitz decay, then the
spectra become flatter due to the contributions from $\eta$ Dalitz
($M\le 0.4$ GeV) and direct $\rho$ decays. At $M\sim 0.78$ GeV the
peak from the direct decay of $\omega$ mesons is visible. Thus,  such
type of 3-dimensional experimental information (or even 4-dimensional
including rapidity) allows to select the contributions from different
channels.

\section{Summary}

Within the framework of the combined resonance-string approach we have
performed a detailed study of dilepton production from $pp$, $pd$ and
$pBe$ collisions at 1.0 -- 4.9 GeV. In this approach the particle
production proceeds through resonances at low energies and by string
formation and decay at high energies. The collisional dynamics is
described by the coupled-channel BUU transport approach. As dilepton
sources we have considered $\pi^0$, $\eta$, $\omega$ and $\Delta$
Dalitz decays, $pn$ bremsstrahlung and direct decays of vector mesons
($\rho$, $\omega$).

We have compared our calculated results to the experimental data of the
DLS Collaboration which provide constraints on the different individual
production channels.  It has been found that the DLS data for the $pp$
and $pd$ collisions can be reasonably well described in our model
whereas for $pBe$ systems (especially at 4.9 GeV) our calculations give
a slightly higher dilepton yield. We have demonstrated the importance
to measure the dileptons from $pp$ and $pd$ (or even $pBe$) collisions
simultaneously since such data can provide constraints on the isospin
dependence of $pp$ and $pn$ interactions, which is important for an
understanding of heavy-ion data.

Dilepton spectra from elementary reactions will be measured in future by
the HADES detector at GSI Darmshtadt with high mass resolution and good
accuracy.  In this respect we have made predictions for the dilepton
invariant mass spectra, transverse momentum and rapidity distributions
for $pp$, $pd$ and $pBe$ at 1.5 and 4.0 GeV.  We showed that proper
cuts in invariant mass for transverse momentum and rapidity spectra
allow to select different dilepton sources and to study, for example,
the $\rho$ meson channel in more detail. We have indicated,
furthermore, that it might be very useful to provide experimentally
multi-dimensional information, e.g. double differential dilepton
spectra $d\sigma/dMdp_T$, in order to investigate the individual
contributions.

\section*{Acknowledgements}
The authors are grateful for valuable discussions with J. Friese,
M. Kagarlis and A. Sibirtsev.


\newpage

\phantom{a}\newpage

\end{document}